\documentclass[aip,preprint,showpacs]{revtex4}
\usepackage{amssymb}
\usepackage{fancyhdr}
\usepackage{dblfloatfix}
\usepackage{array}
\newcolumntype{P}[1]{>{\centering\arraybackslash}p{#1}}

\usepackage{graphicx}

\usepackage[T1]{fontenc}
\usepackage[american]{babel}
\usepackage{color}
\graphicspath{{images/}}
\begin{document}

\title{Complexities in the growth and stabilization of polar phase in the Hf$_{0.5}$Zr$_{0.5}$O$_2$ thin films grown by Pulsed Laser Deposition}
\author{D. Kumar$^{{1},{2}}$}
\thanks{dkumar17@umd.edu}
\affiliation{$^1$CNR-SPIN, Complesso Universitario di Monte S. Angelo, Via Cintia, Naples 80126, Italy; \\$^{2}$Maryland Quantum Materials Center and Department of Physics, University of Maryland, College park, Maryland, 20742, USA;}

\begin{abstract} 
After the discovery of ferroelectricity in HfO$_2$ -- based thin films a decade ago, ferroelectric Hf$_{0.5}$Zr$_{0.5}$O$_2$ (HZO) thin films are frequently being utilized in the CMOS (Complementary Metal-Oxide Semiconductor) and logic devices, thanks to their large remnant polarization, high retention and endurance. A great deal of effort has been made towards understanding the origin of ferroelectricity in epitaxial HZO thin films and controlling the microstructure at the atomic level which governs the ferroelectric phase. Nevertheless, the HZO films still suffer from fundamental questions, such as (1) the vagueness of interfacial mechanisms between HZO, buffer layer and the substrate which controls the polar phase; (2) the nature of the metastable polar phase responsible for the ferroelectricity, be it orthorhombic or rhombohedral; which are poorly understood. Here, we have addressed these issues by employing the {\em in-situ} reflection high energy electron diffraction - assisted pulsed laser deposition and mapping the asymmetrical polar maps on high quality HZO films grown on functional perovskite oxide substrates. The interface between La$_{0.7}$Sr$_{0.3}$MnO$_3$ (LSMO) and the substrate is shown to be quite important, and a slightly rougher interface of the former destabilizes the ferroelectric phase of HZO irrespective of well-controlled growth of the ferroelectric layers. A rhombohedral-like symmetry of HZO unit cell is extracted through the x-ray diffraction asymmetrical polar maps. The ferroelectric measurements on a $\sim$ 7 nm HZO film on STO(001) substrate display a remnant polarization close to 8 $\mu$C/cm$^2$. These results highlight the complexities involved at the atomic scale interface in the binary oxides thin films and can be of importance to the HfO$_2$-based ferroelectric community which is still at its infancy.  


\quad 

\end{abstract}
\maketitle
\newpage
\section{Introduction}
The complex oxide thin films grown by the minute deposition offers many functional properties, such as ferroelectricity, ferromagnetism, colossal magnetoresistance, superconductivity, and so on \cite{Haeni, Deepak1, Deepak2, Deepak3}. Among them, the ferroelectric hafnium oxide-based thin films owing to their seamless compatibility with Si, robust polarization at the nanoscale, possess huge potential for myriad of applications, notably in spintronics and microelectronic device industry \cite{Ramesh, Thomas, Yoong, Mueller}. The research in this area has skyrocketed, specially after the discovery of epitaxial HfO$_2$-based thin films permitting to better understand their underlying mechanisms governing properties \cite{Shimizu}. 
External doping of dopants such as Si, Y, Zr in HfO$_2$ thin films have been recently investigated in details \cite{Shimizu, Lyu1, NM, Lyu2}. Among which, the Zr-substituted HfO$_2$ thin films (Hf$_{0.5}$Zr$_{0.5}$O$_2$ (HZO)) are the most explored ones due to the similar sizes of Hf and Zr (r $\sim$ 160{\em pm}) producing robust polarization at the nanoscale. The presence of ferroelectricity in HfO$_2$-based thin film of thickness as low as 1 nm makes them standout in the ferroelectric community for defying the laws of depolarization field at the nanometric scale \cite{Nature}. On the contrary, the ubiquitous, conventional perovskite ferroelectrics display the {\em{size-effect}} where polarization disappears below a critical film thickness due to the incomplete screening of the surface charges \cite{Ahn}. The reason, {\em{why the ferroelectricity persists down to 1 nm in HfO$_2$ based thin films ?}} is still unclear. Moreover, the very origin of ferroelectricity in these films is still debated to the date. Various possible mechanisms have been held responsible for the stabilization of the ferroelectric phase such as, stress \cite{Park1}, surface energy \cite{Materlik, Park2}, confinement by top electrode \cite{Kim} and so on. \par
The spontaneous polarization observed in HfO$_2$-based thin films is generally ascribed to the metastable polar orthorhombic phase (o, space group {\em{Pca2$_1$}}). However, a slightly higher energetic rhombohedral phase (r, space group {\em{R3m/R3}}) has also been recently reported for HZO thin films grown on SrTiO$_3$ (STO) (001) substrate-buffered with La$_{0.7}$Sr$_{0.3}$MnO$_3$ (LSMO) bottom electrode \cite{NM}. The Density functional theory (DFT) calculations predict that both orthorhombic (space group {{\em Pca2$_1$}}) and rhombohedral (space group {{\em R3, R3m}}) polymorphs fall in the energy window possible to achieve experimentally, although the polar o-phase possess the least energy ({\em i.e.} E({\em Pca2$_1$}) - E({\em P2$_1$/c}) = 64 meV/f.u.) \cite{NM, Materlik}, where {\em P2$_1$/c} represents the ground state monoclinic phase of the bulk compound. The polar ferroelectric phase in the HZO films is often controlled by many external stimuli such as, film thickness, choice of bottom electrode, substrate, and capping layer \cite{window, Estandia, stress, Liu}. The quality of the interface is, therefore, of highest relevance to stabilize the polar phase in the HZO films grown on functional perovskite oxide substrates. The interface between HZO and LSMO has been investigated in detail \cite{Nukala, Tengfei} and the La contents in LSMO have been put forward as one of the significant factors in improving the polarization of HZO films \cite{Estandia}. The use of other metallic bottom electrodes, such as SrRuO$_3$, LaNiO$_3$ strongly diminishes the formation of polar phase in the HZO films. Nevertheless, the interface between LSMO and the underlying substrate, which is of critical importance to the HZO polar phase, has never been discussed and reports are scarce to the best of our knowledge. In this paper, we have addressed these issues by monitoring the growth of HZO thin films by Reflection High Energy Electron Diffraction (RHEED) - assisted Pulsed Laser Deposition (PLD), and scrutinizing the asymmetrical polar maps. Our results show that controlling the layer-by-layer epitaxy of few monolayers of LSMO film interfaced with the substrate is of utmost importance for the stabilization of polar phase in HZO films, besides fulfilling HZO growth criteria. In fact, a slightly incoherent interface of LSMO film with the substrate strongly suppresses the formation of metastable polar phase in HZO film. Moreover, a rhombohedral unit cell of HZO is proposed based on our x-ray diffraction asymmetrical polar maps analysis. The remnant polarization P$_r$ close to 8 $\mu$C/cm$^2$ is obtained for $\sim$ 7 nm HZO film on STO(001) substrate-buffered with LSMO bottom electrode. These observations provide insights into the complexities involved in the stabilization of polar phase in HZO thin films and paves the way for better understanding of robust ferroelectricity in sub-nanometric HZO films.

\section{Experimental} 
The epitaxial bilayers of Hf$_{0.5}$Zr$_{0.5}$O$_2$ (HZO) and La$_{0.7}$Sr$_{0.3}$MnO$_3$ (LSMO) were grown, in a single process, by using Reflection High Energy Electron Diffraction (RHEED)-assisted Pulsed Laser Deposition (PLD), on SrTiO$_3$ (STO) (001) and GdScO$_3$ (GSO) (110) perovskite substrates. The KrF excimer laser with a wavelength of 248 nm was used to ablate the polycrystalline targets of LSMO and HZO placed at a distance of $\sim$ 4 cm from the heater. The LSMO, acting as the bottom electrode, was grown at a laser fluence of $\sim$ 1.2 J/cm$^2$, laser repetition rate of 1-2 Hz, growth temperature and oxygen partial pressure of 730$^\circ$C, and 0.1 mbar, respectively. The thickness of LSMO film was well-monitored by using {\em in-situ} RHEED, and maintained at 20 unit cells. The subsequent HZO layers were grown at a laser fluence of $\sim$ 1.5 J/cm$^2$, laser frequency of 2 Hz, growth temperature and pressure of 800$^\circ$C and 0.1 mbar, respectively. After deposition, the film was cooled down to room temperature at a ramp rate of 10$^\circ$C/min under an oxygen partial pressure of 0.2 mbar. The crystal structure and symmetry analysis of the HZO films were characterized by using a Panalytical X’pert Pro diffractometer. The asymmetrical polar maps were collected by scanning the sample in $\phi$ (azimuth) and $\chi$ (tilt) at a fixed Bragg angle of the detector. The top electrodes (Au, Pt) were deposited using sputtering, and pads of different sizes were processed by lithography. The ferroelectric polarization measurements were performed in the top-bottom configuration using a standard AixACCT TFAnalyser2000 platform.

\section{Results and discussion}
\begin{figure}
\centering
\includegraphics[width = 0.9\textwidth]{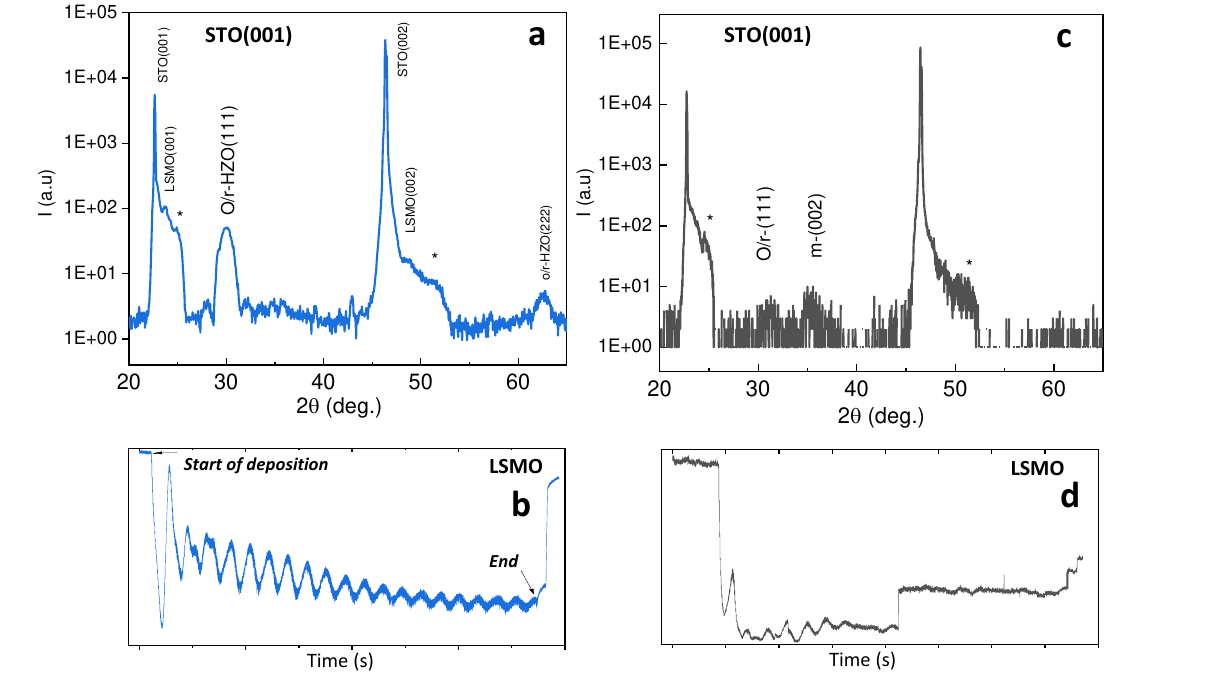}
\caption{{\em (color online) X-ray diffraction $\theta-2\theta$ scans of Hf$_{0.5}$Zr$_{0.5}$O$_2$ (HZO) thin films on STO (001) substrate-buffered with LSMO, for film (a) where LSMO is grown ideally, and (c) for film where LSMO growth is intentionally degraded. Their Reflection High Energy Electron Diffraction (RHEED) specular reflection intensities are also shown, respectively, in (b) and (d). The peaks labeled with asterisk (*) in (a) and (c) are artifacts from the diffractometer.}}
\label{fig 1}
\end{figure}

Fig. \ref{fig 1} shows the nominal $\theta-2\theta$ x-ray diffraction (XRD) scans of two Hf$_{0.5}$Zr$_{0.5}$O$_2$ (HZO) thin films grown on SrTiO$_3$ (STO) (001) substrate-buffered with La$_{0.7}$Sr$_{0.3}$MnO$_3$ (LSMO). The two films differ solely in terms of LSMO growth (i.e., slightly different growth temperatures, laser frequencies etc.) while keeping the HZO growth parameters same for both films. Figs.\ref{fig 1}a,b correspond to the sample where LSMO has been grown perfectly on STO substrate, with quintessential layer-by-layer epitaxy. Fig. 1b clearly shows the Reflection High Energy Electron Diffraction (RHEED) specular reflection intensity oscillations, indicating an epitaxial {\em 2D} layer-by-layer growth of LSMO film on STO substrate. This has a direct influence on the growth of subsequent HZO layers and control over the stabilization of the metastable orthorhombic/rhombohedral - (111) polar phase (hereafter o/r-111 phase) in HZO films, occuring at 2$\theta$ $\sim$ 30$^\circ$ in symmetric $\theta -2\theta$ XRD scan \cite{Qiucheng, stress}.
 This phase has been reported by many research groups to be associated with either orthorhombic ({\em Pca}2$_1$ space group)\cite{stress, Nature}, or rhombohedral ({\em R3m} space group)\cite{NM, Laura, Nukala, Zhang} symmetry of the HZO unit cell and is still at the center of the investigation. We refer to this phase as \textquotedblleft{o/r-111 phase\textquotedblright} in the first part of the manuscript and devote the second part for the exploration of its symmetry nature. Fig. \ref{fig 1}a shows the presence of o/r-111 phase at around 30$^{\circ}$, with the ratio of intensity of o/r-111 peak to the STO (001) {\em i.e.} normalized intensity ({\em I$_{o/r}$/I$_{STO}$} hereafter) close to 0.009, with no signature of monoclinic - (002) phase (m-002) at around 35$^{\circ}$ . On the contrary, the HZO/LSMO heterostructure, where LSMO follows a rough (incoherent) growth on STO (001) as compared to the former film, depicted by a few suppressed specular diffraction oscillations, is shown in Figs.\ref{fig 1}c,d. Here, the growth of LSMO was intentionally degraded by adopting the growth parameters slightly different from those of a standard growth of LSMO film on STO substrate. In particular, the laser fluence was slightly reduced while increasing the substrate temperature, which essentially impedes the perfect layer-by-layer growth of LSMO film on the underlying substrate. Consequently, the o/r-111 phase is significantly reduced by 97 \% ({\em I$_{o/r}$/I$_{STO}$} $\sim$ 0.0003), with an emergence of notable m-002 phase at $\sim$ 35$^\circ$. It is important to mention that, a relatively rougher LSMO film merely follows a growth mode which is different from that of a typical layer-by-layer growth (for instance, step flow growth mode), and does not essentially deteriorates the conductivity of LSMO layers, which was later confirmed.


LSMO typically acts as an epitaxial template for the growth of HZO film, and the strain-state of LSMO is shown to substantially tune the amount of metastable o/r polar phase in HZO film \cite{stress}. A rough LSMO growth, in particular, the initial few layers interfacing the substrate, seems to jeopardize the growth of HZO and consequently the stabilization of o/r-111 phases in the HZO film. This indicates the complexities in the growth of hafnia-based thin films and their reliability on the choice of the bottom layer, as shown by Estand\'ia {\em et. al.}.\cite{Estandia} 

\begin{figure}
\centering
\includegraphics[width = 0.9\textwidth]{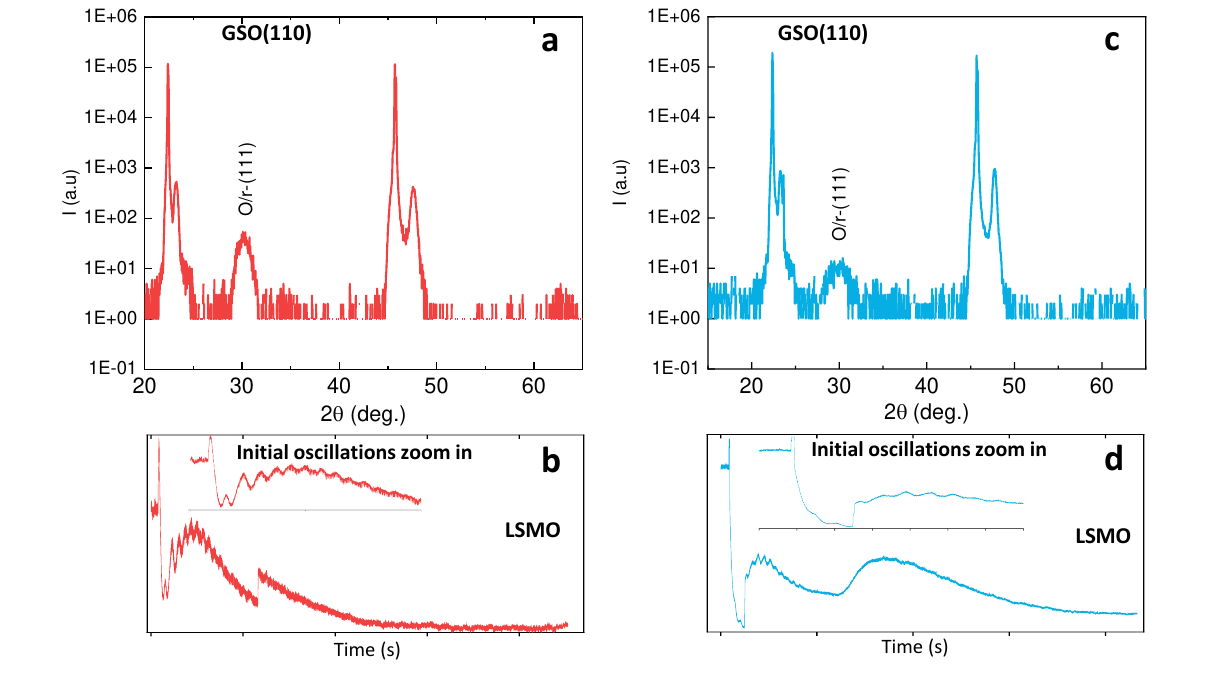}
\caption{{\em (color online) X-ray diffraction $\theta-2\theta$ scans of Hf$_{0.5}$Zr$_{0.5}$O$_2$ (HZO) thin films on GdScO$_3$ (GSO) (110) substrate-buffered with LSMO, for film (a) where LSMO is grown ideally, and (c) for film where LSMO growth is intentionally degraded. Their RHEED specular reflection intensities are shown, respectively, in (b) and (d). Few initial oscillations (specular reflection) are also depicted in the respective insets.}}
\label{fig 2}
\end{figure}

In order to further attest the significance of quality of LSMO film growth and understand the interplay between the interfaces of LSMO/HZO bilayers and LSMO/substrate, we have grown the heterostructure on GdScO$_3$ (GSO) (110) substrate imposing a huge tensile strain ($\sim$ 2.4 \%) on the LSMO layer. Fig. \ref{fig 2} presents a comparison between the XRD symmetric $\theta-2\theta$ scans of two HZO films grown on GSO substrates-buffered with LSMO and the corresponding RHEED specular reflection intensity oscillations of LSMO film on GSO substrate. We have obtained a large o/r-111 phase for HZO film (Fig. \ref{fig 2}a) with {\em I$_{o/r}$/I$_{GSO}$} $\sim$ 0.0004 for the heterostructure where LSMO had been grown smoothly onto the GSO substrate, as shown by initial specular reflection intensity oscillations in Fig. \ref{fig 2}b.
\begin{figure}
\centering
\includegraphics[width = 0.9\textwidth]{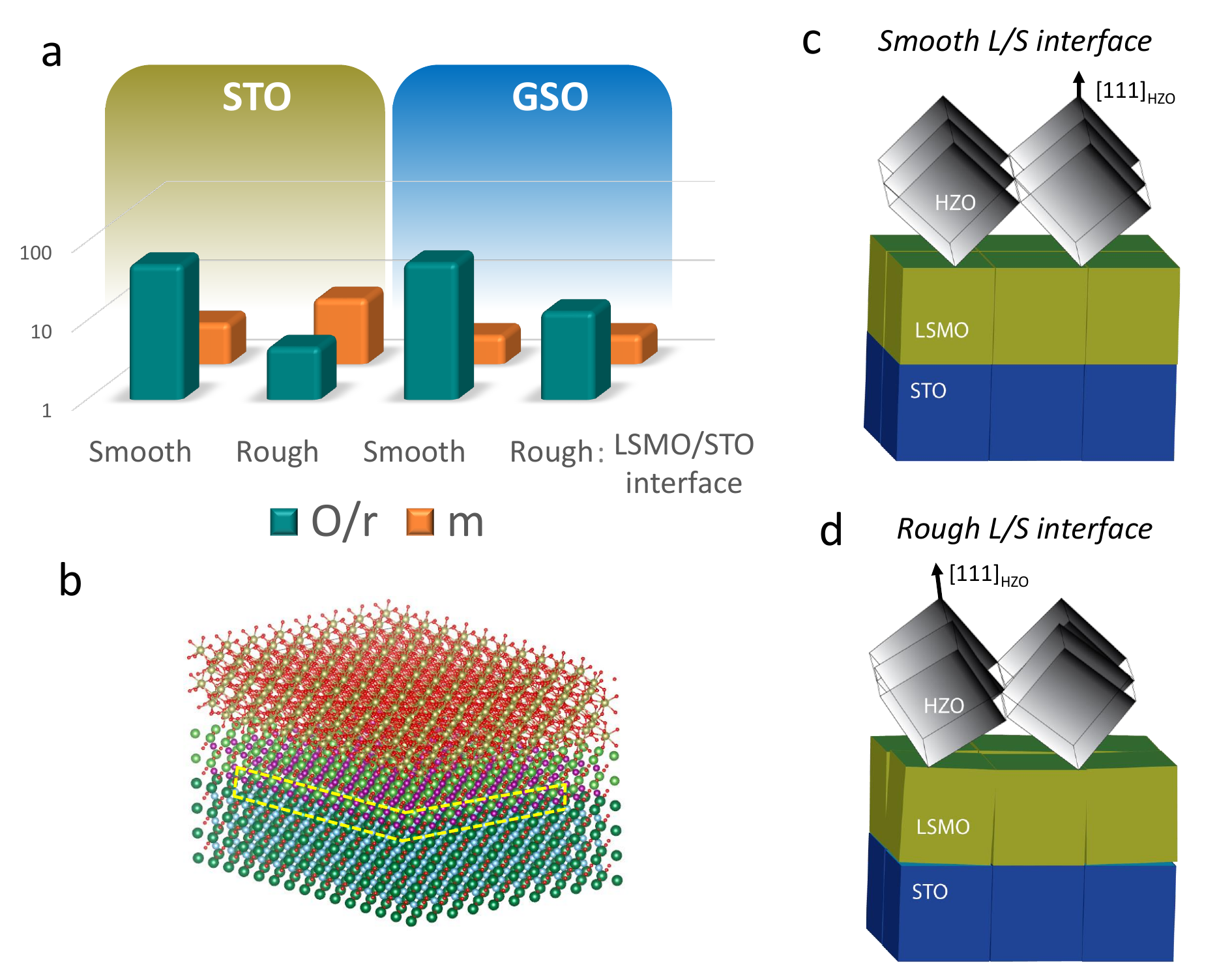}
\caption{{\em (color online) (a) Schematic showing the contribution of each phase (o/r or m), when LSMO is grown coherently (smooth interface), or incoherently (rough interface) on STO substrate. (b) VESTA \cite{Vesta} generated HZO/LSMO/STO heterostructure. The region inside yellow dashed lines emphasizes the interface of LSMO and STO substrate. (c) Schematic representation of an ideal HZO thin film grown on LSMO/STO template, with HZO[111]-axis lying normal to the substrate plane, and (d) an imperfected LSMO growth, leading to a slanted HZO[111]-axis lying slightly off to the normal of the substrate.}}
\label{fig 3}
\end{figure}
 On the other hand, when LSMO film is grown with a slightly rougher interface, shown by a few petite specular reflection oscillations (Fig. \ref{fig 2}d), the {\em I$_{o/r}$/I$_{GSO}$} decreases down to $\sim$ 0.00007 (by 82.5 \%) (Fig. \ref{fig 3}a). We want to emphasize here the importance of the interface between LSMO and the underlying substrate. Although, there have been plenty of work reported focusing on the sensitivity of HZO thin film to the interface it presents with LSMO film, there are none however, to the best of our knowledge, describing the significance of interface between LSMO and the underlying substrate, albeit the latter might seem rather obvious. Moreover, it is worth to note that, despite fewer and suppressed oscillations (Fig. \ref{fig 2}d), the 0/r-111 contribution is relatively higher. This is in contrast to the case of HZO films grown on STO substrate, where a rougher interface strongly diminishes the o/r-111 component. 
\begin{figure}
\centering
\includegraphics[width = 1.0\textwidth]{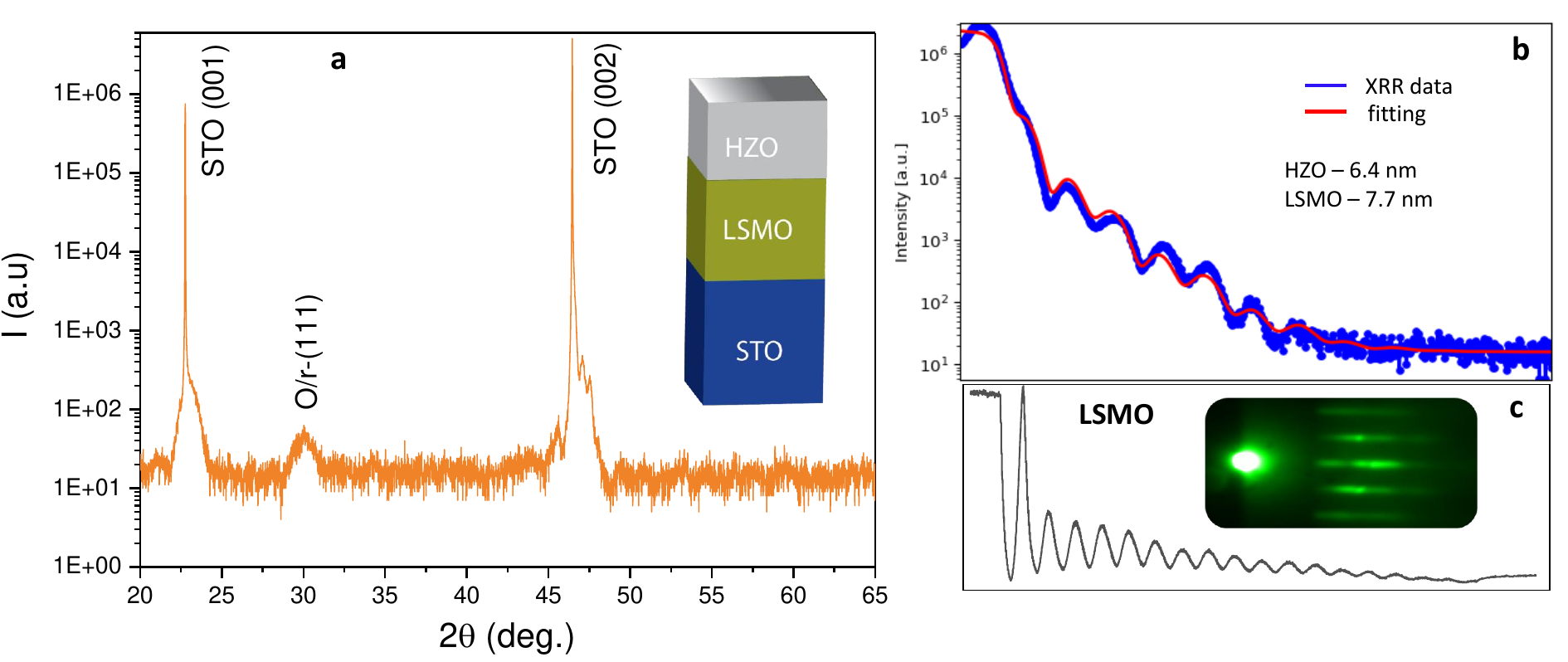}
\caption{{\em (Color online) (a) Nominal x-ray diffraction $\theta-2\theta$ scan of the schematically shown HZO/LSMO/STO(001) heterostructure. (b) Grazing incidence x-ray reflectivity of the aforementioned film. The blue curve is the experimental data, while red curve is the fit using GenX software. (c) RHEED specular reflection intensity for LSMO grown on STO substrate. The inset shows RHEED pattern of LSMO film after deposition.}}
\label{fig 4}
\end{figure}
Therefore, it appears that a decent interface of LSMO with the substrate is a necessity which guarantees the formation of metastable polar phase. This could be due to the fact that a well-controlled, smooth interface between LSMO and the substrate can assist in maintaining the epitaxial strain of LSMO layers intact, thus producing the energetically favorable metastable o/r-111 phase in the strained HZO film, as depicted in fig. \ref{fig 3}c. While, the strain gradient changes drastically in a few layers of LSMO film from the substrate due to rougher LSMO/substrate interface, causing to develop a lower energetic m-002 phase in the relatively defected HZO film (Fig. \ref{fig 3}d).

\begin{figure}
\centering
\includegraphics[width = 0.9\textwidth]{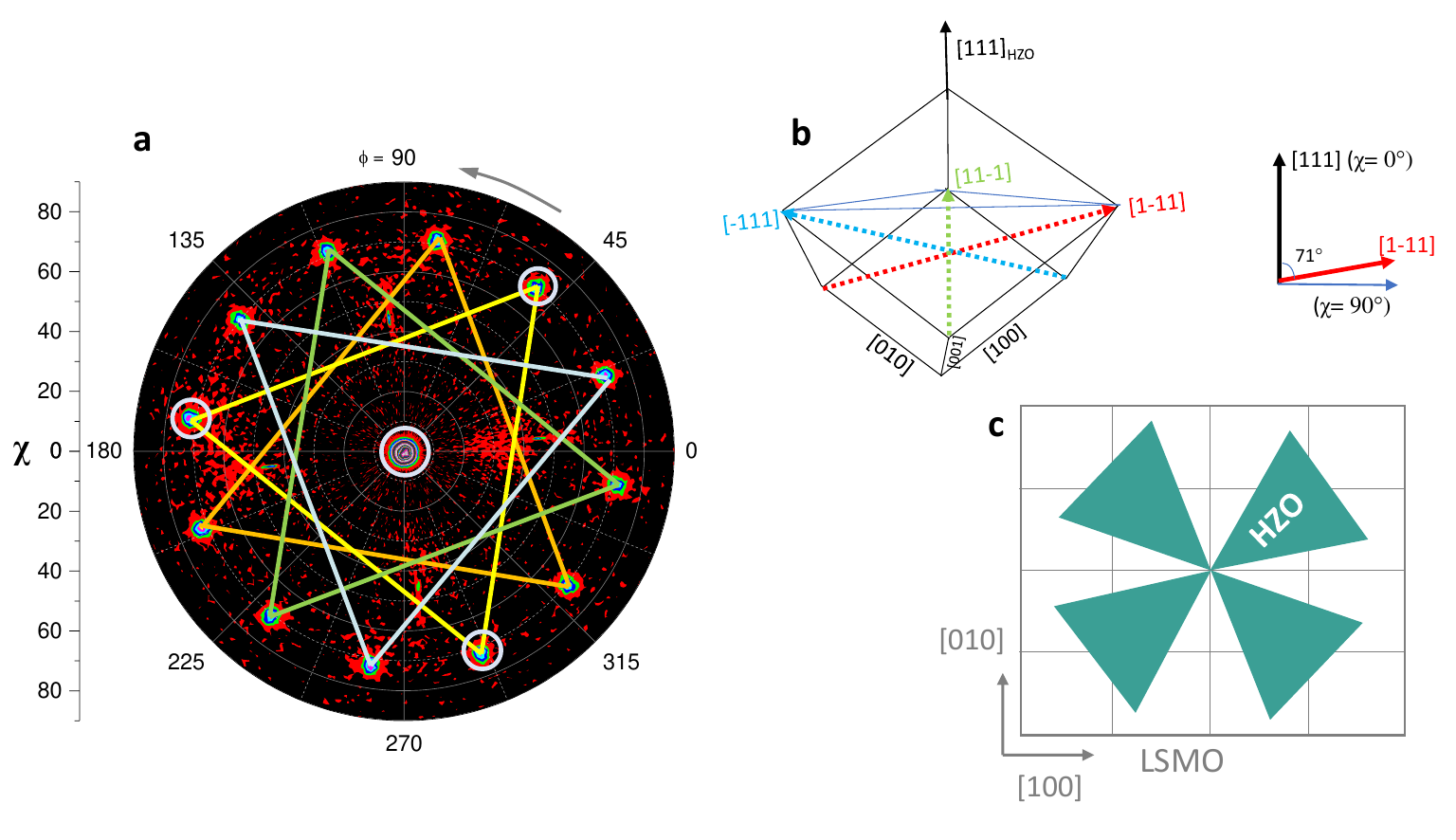}
\caption{{\em (Color online) (a) Pole figures of an HZO film grown on STO(001) substrate-buffered with LSMO, obtained by fixing 2$\theta$ corresponding to the {\em d$_{111}$} and scanning in $\phi$ and $\chi$. The radial axis of the figure represents $\chi$ scale while azimuth shows $\phi$ scale. The four triangles in different colors represent four domains in the HZO film oriented 90$^\circ$ from each other. The white circles show four poles {\em i.e.} (111) (center, at $\chi$ = 0$^\circ$), (-111) (1-11) (11-1) (at $\chi$ $\sim$ 71$^\circ$). (b) Schematic representation of the (111) oriented HZO unit cell, also showing the three poles observed at $\chi$ $\sim$ 71$^\circ$ in (a). (c) Sketch of the epitaxial relationship between HZO and the underlying LSMO/STO template. The top view of the four HZO domains is shown.}}
\label{fig 5}
\end{figure}
After disclosing the role of LSMO/ substrate interface, which could well manipulate the o/r-111 component in HZO films, we performed symmetry analysis of HZO films on STO substrate.
The x-ray diffraction $\theta-2\theta$ and grazing incidence angle reflectivity curve of an HZO film on STO (001) substrate-buffered with LSMO are shown in Figs.\ref{fig 4}a,b, respectively. The $\theta-2\theta$ scan clearly reveals a peak at around 2$\theta$ $\sim$ 30$^\circ$, an indication of o/r-(111) phase and, which is ubiquitously known to be associated with ferroelectricity of the films. However, no monoclinic m-(002) phase at 2$\theta$ $\sim$ 35$^\circ$ can be seen (associated with paraelectricity), indicating the film is free from any parasitic component. 
The low angle x-ray reflectivity was also performed to quantify the thickness of HZO and LSMO films, which were estimated to be around 6.4 nm and 7.7 nm using simulation (GenX), respectively. The growth of LSMO film on STO(001) is derived to be of 2D epitaxial nature, confirmed by the RHEED streak-type pattern (inset of Fig. \ref{fig 4}c) and the specular beam oscillations, as shown in Fig. \ref{fig 4}c. In contrast, the growth of HZO layers on top of LSMO/STO is rather cumbersome and novel, shown by its prevalent (111)-oriented epitaxy on most of the perovskite oxide substrates \cite{stress, window, NM}. Moreover, a high mismatch between HZO and LSMO unit cell produces a rather step-flow-type growth (not shown). \\

\begin{figure}
\centering
\includegraphics[width = 0.9\textwidth]{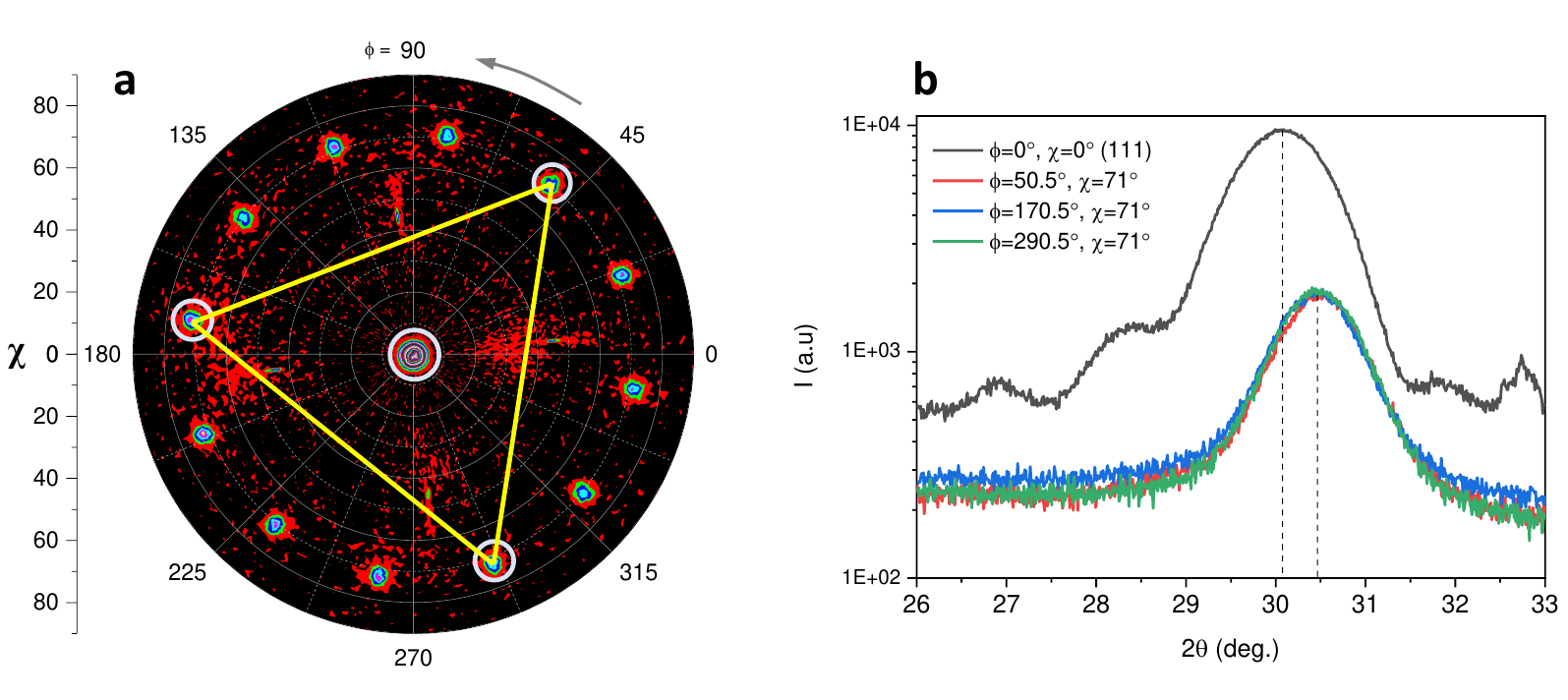}
\caption{{\em (color online) (a) Pole figures of an HZO film grown on STO(001) substrate-buffered with LSMO electrode. The yellow triangle with three poles at its corners represent a single domain. (b) XRD $\theta-2\theta$ scans around each of the encircles poles in (a).}}
\label{fig 6}
\end{figure}

The representative \{111\} pole figure (set of $\phi-\chi$ scans) obtained around the out-of-plane axis of HZO is shown in Fig. \ref{fig 5}a. For this, the 2$\theta$ was fixed at $\sim$ 30$^\circ$ (estimated from the o/r-111 peak in Fig. \ref{fig 4}a) and the sample was scanned in $\phi$ (0 to 360$^\circ$) and $\chi$ (o to 90$^\circ$). The presence of 12 poles at $\chi$ $\sim$ 71$^\circ$ is a result of four domains oriented 90$^\circ$ from each other with respect to the film normal (as shown in Figs. \ref{fig 5}a,c), where in each domain contributes to three poles. Fig. \ref{fig 5}b presents a schematic of HZO unit cell grown on LSMO/STO template with (111)-axis lying along the out-of-plane direction, essentially appearing at $\chi$ = 0$^\circ$ in Fig. \ref{fig 5}a. Moreover, the other three poles {\em i.e.} (-111), (1-11) and (11-1) are also shown, appearing at $\chi$ $\sim$ 71$^\circ$ and separated by 120$^\circ$ in $\phi$.\par
To further analyze the lattice symmetry of HZO unit cell, normal $\theta-2\theta$ scans were performed around each of the aforementioned poles, and shown in Fig. \ref{fig 6}. For simplicity, we have chosen a single domain (shown in Fig. \ref{fig 6}a) and performed $\theta-2\theta$ scans around its four poles (encircled in white), which are depicted in Fig. \ref{fig 6}b. It is evident that, the Bragg angles corresponding to all the poles at $\chi$ $\sim$ 71$^\circ$ are equal, and larger than the pole at $\chi$ = 0$^\circ$. This indicates a larger{\em d}-spacing along the out-of-plane (111)-axis compared to other three inclined axes, following {\em d$_{[111]}$} (2.96 \AA) $>$ {\em d$_{[-111]}$} (2.93 \AA)={\em d$_{[1-11]}$} ={\em d$_{[11-1]}$}. The 3:1 multiplicity in the d-spacings of \{111\} planes, as observed here, is generally a signature of the rhombohedral symmetry. In contrast, the other polymorphs such as, orthorhombic, cubic, tetragonal have the atomic planar spacing along the {\em out-of-plane} direction ({\em{d$_{111}$}}) similar to that along the [-111], [1-11] and [11-1] directions. Therefore, these measurements suggest a rhombohedral symmetry of the HZO cell, which is experimentally less probable to stabilize due to its higher energy as compared to the metastable orthorhombic phase \cite{Zhang}. \par

\rightline{Now, we resort to the ferroelectric switching measurements on the twin} of this sample, which was especially grown for the electrical measurements. Before the top electrode (Au) fabrication process, the sample was first scanned under XRD to check for the desired metastable polar phase, a prerequisite, and was observed to contain the r-111 phase. 
\begin{figure}
\centering 
\includegraphics[width = 1.0\textwidth]{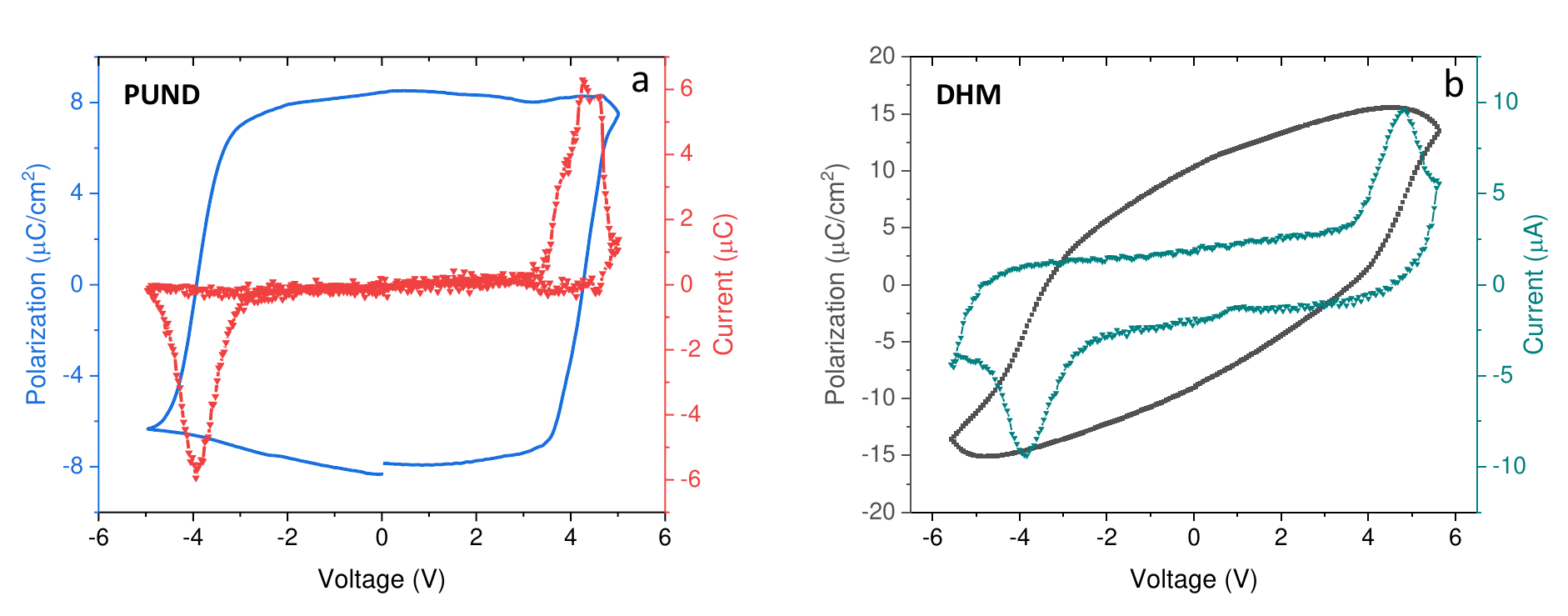}
\caption{{\em (color online) The ferroelectric polarization measurements of a $\sim$ 7 nm HZO thin film grown on LSMO-buffered STO(001) substrate. The (a) Positive Up Negative Down (PUND) and (b) Dynamic Hysteresis Mode (DHM) measurements with in-built leakage compensation, performed at 1 kHz and at 300 K are shown.}}
\label{fig 7}
\end{figure}
Fig. \ref{fig 7} shows the Positive Up Negative Down (PUND) and Dynamic Hysteresis Mode (DHM) measurements with the in-built leakage compensation of a $\sim$7 nm HZO film on STO(001) substrate-buffered with the LSMO bottom electrode, performed at 1kHz. The dynamic leakage current compensation effectively subtracts any leakage current due to the flow of electrons in the sample, specifically at the higher frequencies where current due to the ferroelectric switching dominates over the leakage current \cite{ignasi}. From PUND measurements, a proper switching of the ferroelectric domains at the coercive voltage (V$_c$) $\sim$ 4V and the remnant polarization (P$_r$) $\sim$ 8 $\mu$C/cm$^2$ are obtained. The DHM expectedly incorporates the background lossy contribution, and thus, overestimates the estimated P$_r$ $\sim$ 10 $\mu$C/cm$^2$. Our observed value of P$_r$ is slightly smaller than the value reported by other works for the similar HZO films \cite{NM, Nukala}, but comparable to some, as in ref. \cite{Estandia}. The reasons for such reduction in the ferroelectric polarization could be many fold. Firstly, as it is well-known, the ferroelectric polarization typically depends on the size/area of the top electrodes exploited to realize the capacitor, besides film thickness and epitaxial strain induced by using different perovskite substrates of different mismatches. In our work, we have used relatively larger Au electrodes of 30 x 150 $\mu$m$^2$ in size, which can significantly hinder the observation of exact ferroelectric polarization of HZO film. Secondly, it could be related to our observed crystal symmetry of the HZO unit cell. As it has been recently reported by Y. Wei et. al. \cite{NM}, an extensively elongated HZO unit cell along the {\em out-of-plane} direction (d$_{111}$ > 3.2 \AA) is highly polar, with P$_r$ higher than 15 $\mu$C/cm$^2$. However, the P$_r$ plummets rapidly as the {\em out-of-plane} decreases due to smaller in-plane compression. The HZO films in this work typically have d$_{111}$ < 2.97 \AA, falling in the vicinity of reduced polarization.

\section{Conclusions}
In summary, we have grown high quality sub-nanometer HZO thin films on the perovskite oxide substrates. The interface between the bilayers is shown to be critical to realize the parasitic-component free, polar HZO thin films. A slightly incoherent interface between substrate and the bottom electrode destabilizes the ferroelectric polar phase in HZO thin films. A rhombohedral unit cell is extracted for a 7 nm HZO film grown on STO(001) substrate by employing the asymmetrical polar maps, although local microscopy would be needed to get further insights into the structure. The ferroelectric-switching measurements on the film strongly suggest the polar nature of the films, albeit producing smaller polarization most probably due to the large area of the realized pads inducing large leakage. 



\end{document}